\documentclass[12pt]{iopart}

\usepackage{graphicx}

\usepackage{iopams}  
\usepackage{textgreek}

\begin{document}

\title[]{Combined synchrotron X-ray diffraction and NV diamond magnetic microscopy measurements at high pressure}


\author{Lo\"ic Toraille$^{1,2}$, 
Antoine Hilberer$^1$, 
Thomas Plisson$^2$, 
Margarita Lesik$^{1}$\footnote{Current position: Galatea Laboratory, STI/IMT, Ecole Polytechnique F\'ed\'erale  de Lausanne (EPFL), Neuch\^atel, Switzerland.},
Mayeul Chipaux$^{1}$\footnote{Current position: Institute of Physics, Ecole Polytechnique F\'ed\'erale de Lausanne (EPFL), CH-1015 Lausanne, Switzerland.},
Baptiste Vindolet$^1$, 
Charles P\'epin$^2$, 
Florent Occelli$^2$, 
Martin Schmidt$^1$, 
Thierry Debuisschert$^3$, 
Nicolas Guignot$^4$, 
Jean-Paul Iti\'e$^4$, 
Paul Loubeyre$^{2,5}$ and Jean-Fran\c cois Roch$^1$}

\address{$^1$Universit\'e Paris-Saclay, CNRS, ENS Paris-Saclay, CentraleSupelec, LuMIn, 
 91190 Gif-sur-Yvette, France\\
$^2$CEA, DAM, DIF, 91297 Arpajon, France;\\
$^3$Thales Research \& Technology, 1 avenue Augustin Fresnel, 91767 Palaiseau cedex, France\\
$^4$Synchrotron SOLEIL, L'Orme des Merisiers St.Aubin, BP48, 91192 Gif-sur-Yvette, France\\
$^5$Universit\'e Paris-Saclay, CEA, Laboratoire Mati\`ere sous conditions extr\^emes, 91680 Bruy\`eres le Chatel, France}
\ead{jean-francois.roch@ens-paris-saclay.fr}

\vspace{10pt}

\begin{abstract}
We report the possibility to  simultaneously perform wide-field nitrogen-vacancy (NV) diamond magnetic microscopy and synchrotron X-ray diffraction (XRD) measurements at high pressure. NV color centers are created on the culet of a diamond anvil which is integrated  in a  diamond anvil cell  for static compression of the sample. The optically detected spin resonance of the NV centers is used to map the  stray magnetic field produced by the sample magnetization.
 Using this combined scheme, the magnetic and structural behaviors can be simultaneously  measured. As a proof-of-principle, we record the correlated  \textalpha-Fe to \textepsilon-Fe structural and magnetic transitions of iron that occur here between 15 and 20~GPa at 300~K. 
\end{abstract}

\vspace{2pc}

%
%
%

\newpage

Pressure can be used as a tuning parameter to disclose relations between magnetic, structural and  electronic properties of a solid-state system. Much detailed measurements with far reaching applications have been made thanks to   the development of the diamond anvil cell (DAC)~\cite{mao2018} which is an efficient and practical tool to generate hydrostatic static compression up to extreme pressures in the megabar range. 
The development of high-pressure physics is also tightly linked to the improvement of synchrotron radiation sources~\cite{shen2017} that now feature focal spots in the micrometer range along with high brightness in the 20-40 keV range. 
Due to the transparency of diamond to X-rays, a broad panel of X-ray synchrotron-based methods can be applied for high-pressure studies as in situ probes of the materials inside the DAC. As an example, complex magnetic ordering can be resolved using X-ray magnetic circular dichroism (XMCD) which provides     element-specific probing of magnetic moments~\cite{itie2016}. 
Other magnetic measurement techniques have been developed at high pressure but remain challenging due to the constraints of their implementation in the   DAC~\cite{eremets96}. We report here that wide-field nitrogen-vacancy (NV) diamond magnetic microscopy based on the imaging of an ensemble of NV centers~\cite{levine2019} can be  implemented on a synchrotron X-ray diffraction (XRD) beamline so as to address both the unit cell structure and volume of the compressed sample inside the DAC. The combination of these two diagnostics  allowed us to measure the magnetic properties of the   sample  and to simultaneously determine its cristallographic structure by recording the   XRD pattern.
The technique uses a specifically engineered diamond anvil and a compact optical setup which is slid on and off  the XRD bench.  

The negatively-charged NV$^-$ center  is a point defect in diamond which can be engineered as an atomic-like quantum system~\cite{luhmann2018}. It behaves as a magnetic quantum sensor due to the existence of an electron spin $S=1$ associated with a bright photoluminescence (PL)~\cite{doherty2013}. In absence of any magnetic field, this spin structure introduces discrete energy states inside the diamond bandgap. These states have different PL intensities due to their interaction with additional metastable levels. While optically pumping the NV centers with green laser light and simultaneously driving the electronic spin with microwave (MW) tuned at the resonance frequency of 2.87~GHz between the $m_S =0$ (high-PL) and $m_S =\pm 1$ (low-PL) states, the optically detected magnetic resonance (ODMR) signal appears as a decrease of the PL intensity. When a magnetic field is applied to the NV center, the Zeeman effect removes the degeneracy between the $m_S=\pm 1$ states and the resonance ODMR signal is thus split in two components. The  difference between these two resonance frequencies gives access to the projection of the magnetic field on the quantization axis which is defined by the N-to-V direction of the NV center~\cite{taylor2008}.  When collecting the PL from an ensemble of NV centers, the NV-ODMR spectrum consists in the cumulative response of many defects. Since single-crystal diamond only allows four different cristallographic orientations $a$, $b$, $c$ or $d$  for the NV centers, the signal associated to an ensemble of NV centers is   the superposition of the PL emitted by these four families of NV centers~\cite{lai2009}. The NV-ODMR signal of each family is then associated with a given projection of the magnetic field vector in the $\langle 111 \rangle$ set of equivalent symmetry directions. The three vector components of the magnetic field can then be retrieved from these four correlated measurements~\cite{chipaux2015a}. This technique can be extended to a layer of NV centers overlapping a magnetic structure in order to map the full vector magnetic field created by the sample magnetization. 

The quantitative mapping of the magnetic field can then be modeled to determine the magnetization  of micrometer sized magnetic structures~\cite{toraille2018,mccoey2019}. Ensemble-NV sensors  have been applied in a number of applications, including the detection of the magnetic properties of samples in a diamond anvil cell at pressures up to 30 GPa~\cite{hsieh2019,lesik2019,yip2019}. 

\begin{figure}
 \includegraphics[width=0.8\textwidth]{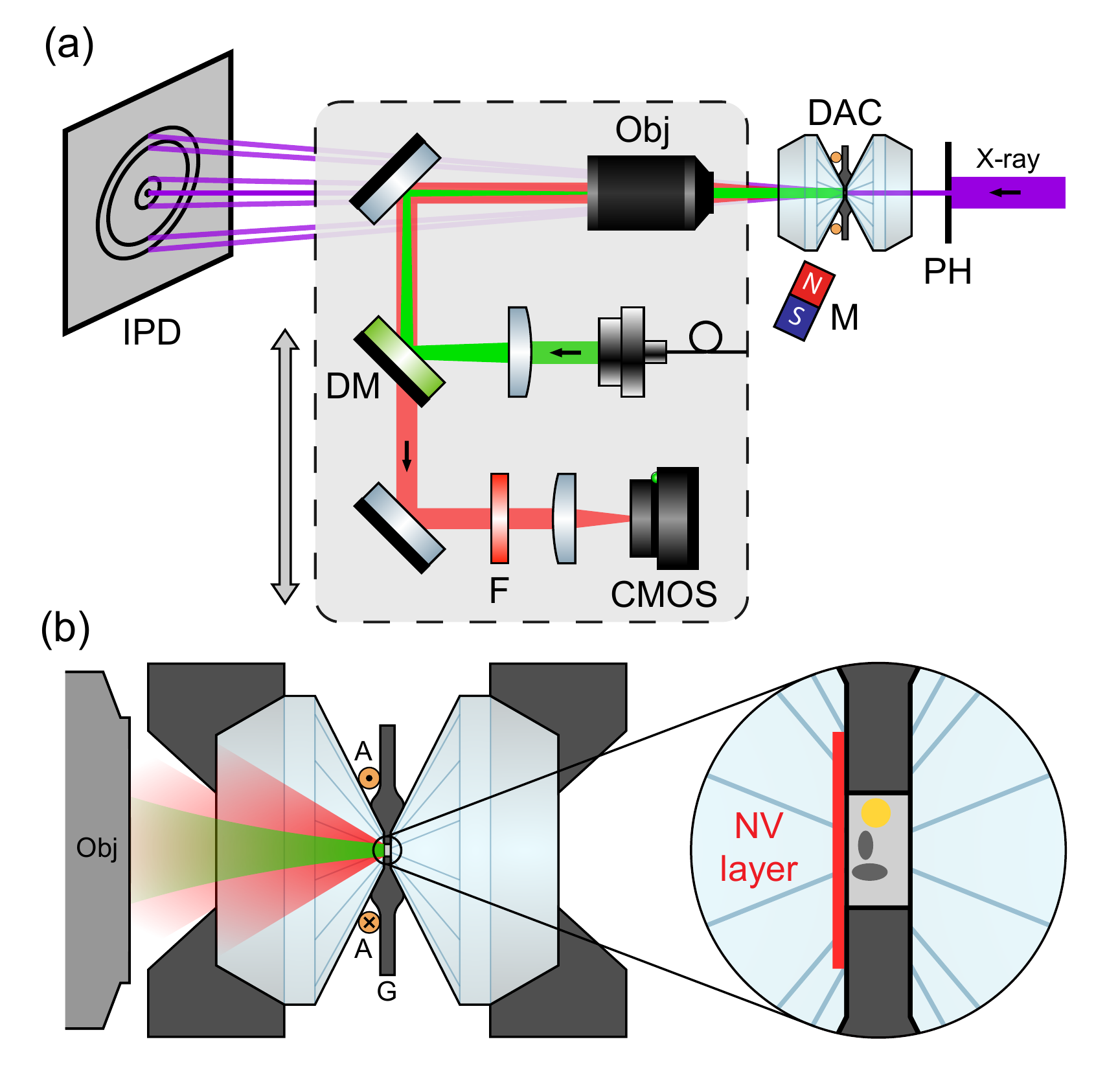}
 \caption{\textbf{(a)} Experimental setup used to sequentially perform  NV-based magnetic measurement and XRD measurement in a DAC. The DAC is composed of type IIas Almax-Boehler anvils with 300~\textmu m-diameter culets. A layer of NV centers was created on the tip of one of the anvils using nitrogen ion implantation. In the wide-field NV diamond magnetic microscope, a microscope objective (Obj, Nikon $\times40$ and numerical aperture of 0.4) focuses a continuous-wave  laser beam at 532 nm wavelength on the NV layer for optical pumping. 
 The red PL emitted by the    NV center layer is   collected by the  same objective, then spectrally filtered from the  laser excitation
  using a dichroic mirror (DM) and a long-pass optical filter (F), and finally imaged on a CMOS camera. When the NV microscope is removed, the DAC is illuminated by the X-ray beam whose size is  limited by a pinhole (PH) of 50 \textmu m diameter  to avoid parasitic diffraction  from the DAC gasket.  
   The XRD pattern is then recorded on an image plate detector (IPD).  The permanent magnet located near the DAC (M)  is used to induce the magnetization of the iron samples and  to discriminate between the four $\langle 111 \rangle$ orientations of the NV centers. 
 \textbf{(b)} Implementation of ODMR in the DAC chamber delimited by the tips of the two anvils and a rhenium  gasket (G). A single loop of insulated copper wire shown in orange is wrapped around the anvil hosting the NV centers as the antenna (A) which provides  the MW excitation for the electron  spin resonance. 
 A slit is cut in the  gasket    to prevent MW screening by avoiding the onset of  eddy currents induced in the rhenium metallic foil. The inset shows a close-up  of the DAC chamber  filled with NH$_3$BH$_3$ as the pressure transmitting medium. Micrometer-size pieces of iron (in grey) and gold (in yellow) are deposited on the tip of the NV-doped anvil. The red line shows the implanted NV layer in the culet of the anvil with a concentration of about $10^4$ defects/\textmu m$^2$ and located at a depth of 20 nm below the surface. }
 \label{figDAC}
 \end{figure}

Elaborating on our work on NV magnetic measurement in a DAC relying on the activation of a layer of NV centers created in the culet of a diamond anvil~\cite{lesik2019}, we built a compact and portable NV diamond  magnetic microscope that ensured its compatibility with a XRD measurement on a synchrotron beamline (figure~\ref{figDAC}.a). This optical microscope, which is used to detect the magnetic field created by the sample,  is constructed so that it can be slid on and off in front of the DAC which is settled on the XRD bench.
The experiment was then performed at the PSICH\'E  beamline of the SOLEIL synchrotron facility (Gif-sur-Yvette, France).  Powder XRD patterns were collected using a monochromatic X-ray beam at 0.3738\,\AA \;  wavelength and a Pilatus CdTe image plate detector.   During the experiment performed at room temperature, we alternatively aligned the NV magnetometer with the DAC and removed it to free the way towards the X-ray detector used to record the XRD pattern. 
In the conditions of the experiment, we did not observe any modification of the NV centers luminescence properties that could be induced by the exposure to the X-ray beam, such as the potential photoionization of the negatively charged state NV$^-$ into the neutral charged state NV$^0$~\cite{aslam2013}.  

To demonstrate the complementarity between XRD and NV diamond magnetic microscopy methods we chose the \textalpha-Fe$ \, \leftrightarrow \,  $\textepsilon-Fe~phase transition of iron~\cite{Bassett1987} as  a testbed example. The structural transition between the body-centered-cubic (bcc) \textalpha~phase and the hexagonal close-packed (hcp) \textepsilon~phase is associated with a magnetic transition between the ferromagnetic state of \textalpha-Fe and the mainly non-magnetic state of \textepsilon-Fe. Both transitions have been extensively studied due to their implications in iron-based materials and their relation with magnetism and Earth sciences~\cite{mathon2004,baudelet2005,monza2011,dewaele2015a,wei2017,lebert2019}. 

As shown in figure~\ref{figDAC}.b, the DAC is made of non-magnetic material and a rhenium gasket is used to make the sample chamber.  
At the time of the experimental run, the spot size of the X-ray beam  available at the PSICH\'E  beamline had a diameter of  about 80~\textmu m.  In order to maximize the XRD signal we then had to use a sample consisting of several iron beads of 99.5\% purity which were packed together. As a consequence, the inhomogeneous magnetization of those irregularly packed samples of iron  created a complex magnetic field distribution 
that could not be easily inverted to determine   the magnetization texture~\cite{vervelidou2017}. Note that this experimental feature is a significant difference with our previous results of  Ref.~\cite{lesik2019} where  we measured the  magnetic field created  by an iron bead of 5 \textmu m size. The magnetic response of this single iron bead could then be modeled as a simple magnetic dipole, leading to a  quantitative measurement   of the  iron magnetization decrease along the  \textalpha-Fe to \textepsilon-Fe transition. 
 
The iron beads were  embedded in NH$_3$BH$_3$   for pressure transmission in the DAC.  A gold bead of about 5~{\textmu}m diameter was added in the DAC chamber to serve as a pressure gauge using the calibrated equation of state of gold~\cite{takemura2008} to fit the gold unit cell volume. Although the NH$_3$BH$_3$ salt  induced a distortion of the samples optical microscopy images   that could be recorded through the diamond anvil cell, it had no influence on the quality of the PL images.  Moreover  some   properties of this solid-state pressure transmitting medium  were useful: It is a weak scatterer with a large unit cell and, as shown below, it is not perturbing the iron XRD pattern. It also enables to minimize the gasket deformation which helps to preserve the integrity of the slit made in the metallic gasket for the propagation of the MW excitation to the NV ensemble~\cite{lesik2019}. 
 A bias magnetic field of approximately 9~mT was applied using a permanent magnet oriented close to the DAC. This external magnetic field has two features: it magnetizes the iron samples and it creates a significant splitting of the eight NV-ODMR resonances associated to the four $\langle 111 \rangle$ axes of the NV center~\cite{toraille2018}. Using a membrane actuated DAC,  we progressively increased the pressure  up to 32.7~GPa. At each pressure step we successively recorded the XRD pattern and the NV-ODMR optical images.    

\begin{figure}
 \includegraphics[width=\textwidth]{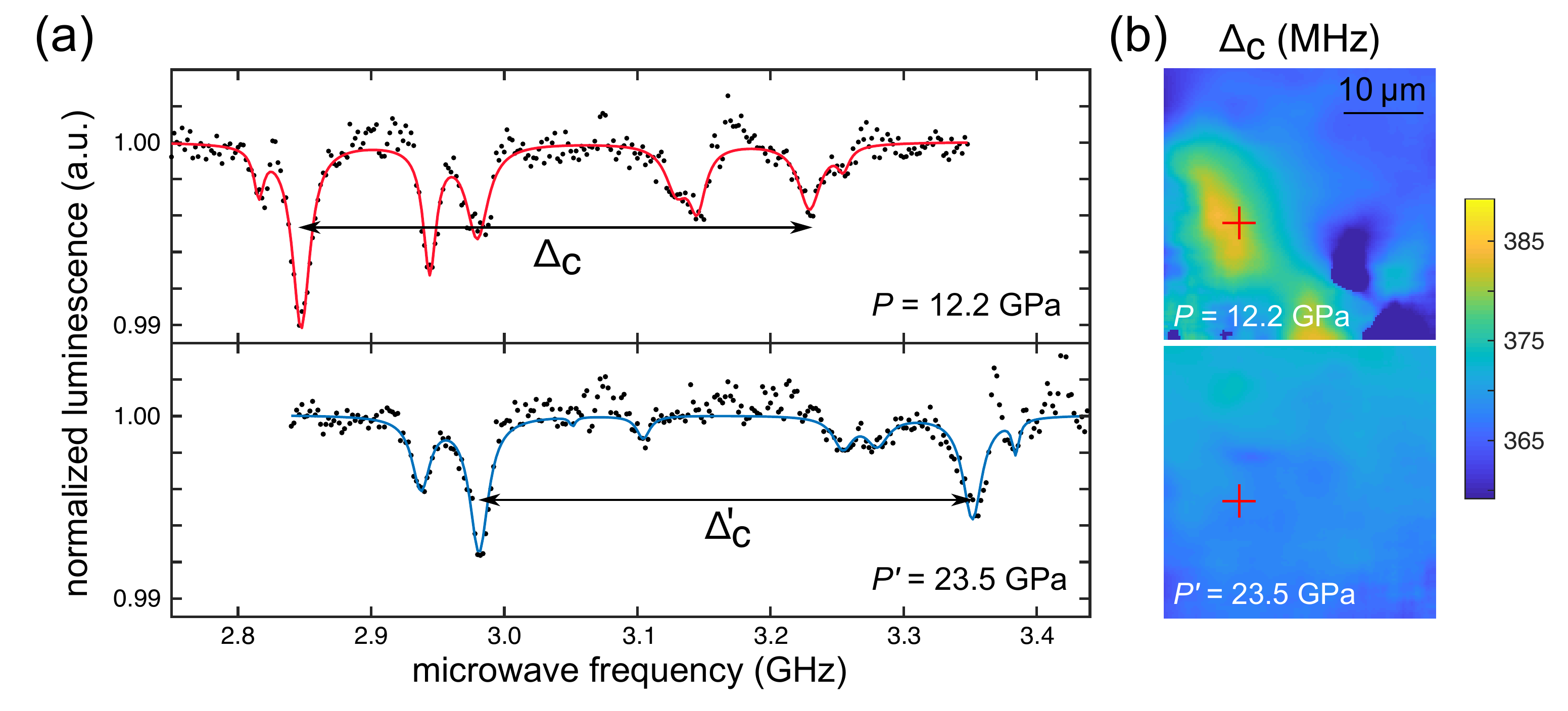}
 \caption{\textbf{(a)} Reconstruction of NV-ODMR spectra for the ensemble of NV centers corresponding to the pixel indicated by the red cross in panel (b), at pressures of $P=12.2\,{\rm GPa}$ (upper curve) and $P^\prime = 23.5\,{\rm GPa}$ (lower curve). 
 The PL intensity is normalized to its mean value taken outside the spin resonance. 
 A bias magnetic field of around 9~mT is used to split the eight resonances associated with the four crystallographic families of NV centers, and to induce the sample magnetization. The splitting from the family exhibiting the highest ODMR contrast is $\Delta_{\mathrm{C}}= 381.9\,{\rm MHz}$ at  $12.2\,{\rm GPa}$ and  decreases to $\Delta^\prime_{\mathrm{C}}= 370.6\,{\rm MHz}$ at $23.5\,{\rm GPa}$. The general shift of the NV-ODMR spectrum is due to the pressure increase. The processed data is shown in black while the red and blue solid lines  are numerical fits using a set of Lorentzian functions. For increased robustness, only the data associated with the fitting of the family $c$ is considered for the rest of the analysis. \textbf{(b)} Corresponding maps  of $\Delta_{\mathrm{C}}$ evaluated above the iron samples. }
 \label{figESR}
 \end{figure}

Figure~\ref{figESR} shows an example of results obtained at a pressure of 12.2~GPa and 23.5~GPa with the NV diamond magnetic microscope. Each pixel of the camera recorded the luminescence of an optically pumped ensemble of NV centers. By sweeping the MW frequency applied to the NV centers we then measured the full ODMR spectrum for each pixel of the image (see figure~\ref{figESR}.a). Compared to the results of Ref.~\cite{lesik2019} significant noise of still unclear origin led to a smaller ODMR contrast which interfered with the processing of the data taken during this test run. To insure the validity of our results, we decided to only select a single set of resonance peaks associated to the orientation $c$ of the NV centers which kept the highest ODMR contrast over the sample area and throughout the whole experiment run. 
Although we performed a least-square minimizing multi-Lorentzian fit of the full NV-ODMR spectrum for added robustness,  we only retrieved the data which are associated with the splitting value $\Delta_{\mathrm{C}}$ of the selected NV center family. We then reconstructed maps of this single splitting in the vicinity of the sample as shown in figure~\ref{figESR}.b. Clear local variations of $\Delta_{\mathrm{C}}$ can be observed with an increase close to the iron sample that reveals the  \textalpha-Fe magnetization.

\begin{figure}
 \includegraphics[width=0.75\textwidth]{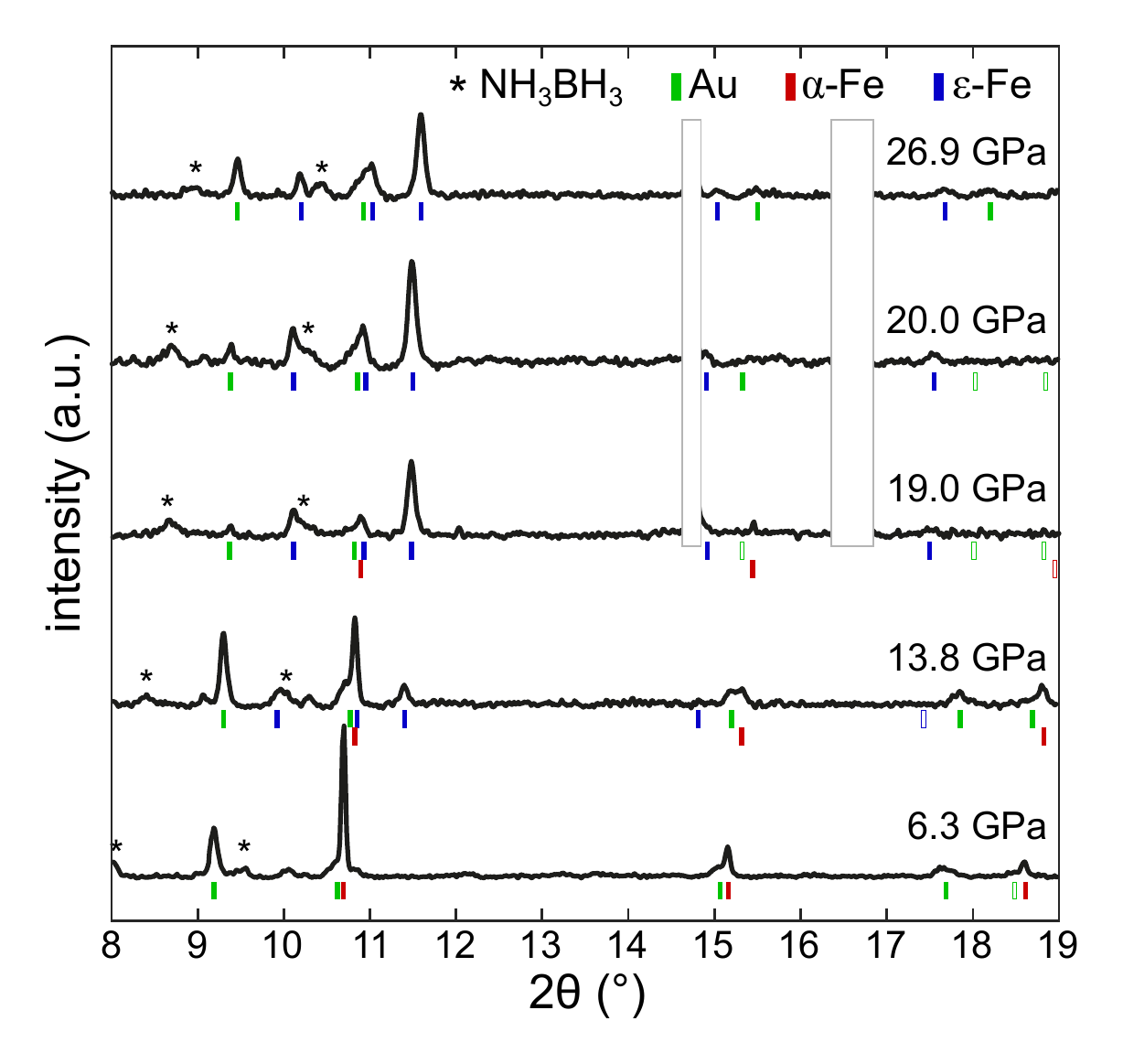}
 \caption{\textbf{(a)} X-ray diffractograms recorded for increasing pressure steps. The data are displayed with a vertical  offset for clarity. At 6.3~GPa, only Au and \textalpha-Fe diffraction peaks appear. At 13.8~GPa, the intensities of the peaks corresponding to \textalpha-Fe have decreased while \textepsilon-Fe peaks can be clearly observed. Disappearance of the last traces of \textalpha-Fe is observed between 19.0  and 20.0~GPa. From 20.0  up to 26.9~GPa, the diffraction signal only shows the presence of Au, NH$_3$BH$_3$ and \textepsilon-Fe. The intensity of the Au diffraction peaks is evolving as the relative positions of the Au bead and the Fe sample change under the pressure induced deformation of the gasket. The  diffraction peaks indicated by the stars are associated with the NH$_3$BH$_3$ salt. The most intense NH$_3$BH$_3$ diffraction peaks are found below 8$^{\circ}$. Additional peaks overlaid with white squares are caused by the diffraction of the permanent magnet used to split the spin resonance of the four NV families and to induce the  magnetization of the iron sample. We observe no angular displacement of these parasitic peaks with pressure changes. Due to the  size of the X-ray beam smaller than the diameter of  the hole in the DAC  gasket, no parasitic peaks associated to  rhenium  are observed. Filled markers indicate discernible peaks, while hollow markers indicate theoretical positions for the peaks which remain hidden in the noise. }
 \label{figXRD}
 \end{figure}

The XRD pattern acquired for each pressure step can be used to follow the \textalpha-Fe$ \, \leftrightarrow \,  $\textepsilon-Fe   structural transition.
 Selected diffractograms, taken before, during, at the end and after this   transition  are shown in figure~\ref{figXRD}.
  Using the Fe and Au equations of state,  we can assign each sharp Bragg peak to a specific crystallographic structure of each element. The diffraction peaks of \textalpha-Fe, \textepsilon-Fe and Au are clearly identified. Two weak peaks associated to NH$_3$BH$_3$  are also observed
but due to its large unit cell the most intense peaks of this compound appear at low diffraction angles, hence outside the region of interest for the transition in iron. 
 The first fraction of the \textepsilon~phase appears at 13.1~GPa while the last fraction of the \textalpha~phase disappears at 20.0~GPa.
  At the end of the experiment run, the pressure was   brought back down to its ambient value. 
 We then observed the inverse transition  with a hysteresis behavior; the \textalpha~phase started to appear again at 12.9~GPa and the \textepsilon~phase completely disappeared at 6~GPa. These values are consistent with those reported in the literature~\cite{monza2011,dewaele2015a}. Finally no chemical reaction between the iron sample and the NH$_3$BH$_3$ salt could be detected. 

\begin{figure}
\includegraphics[width=\textwidth]{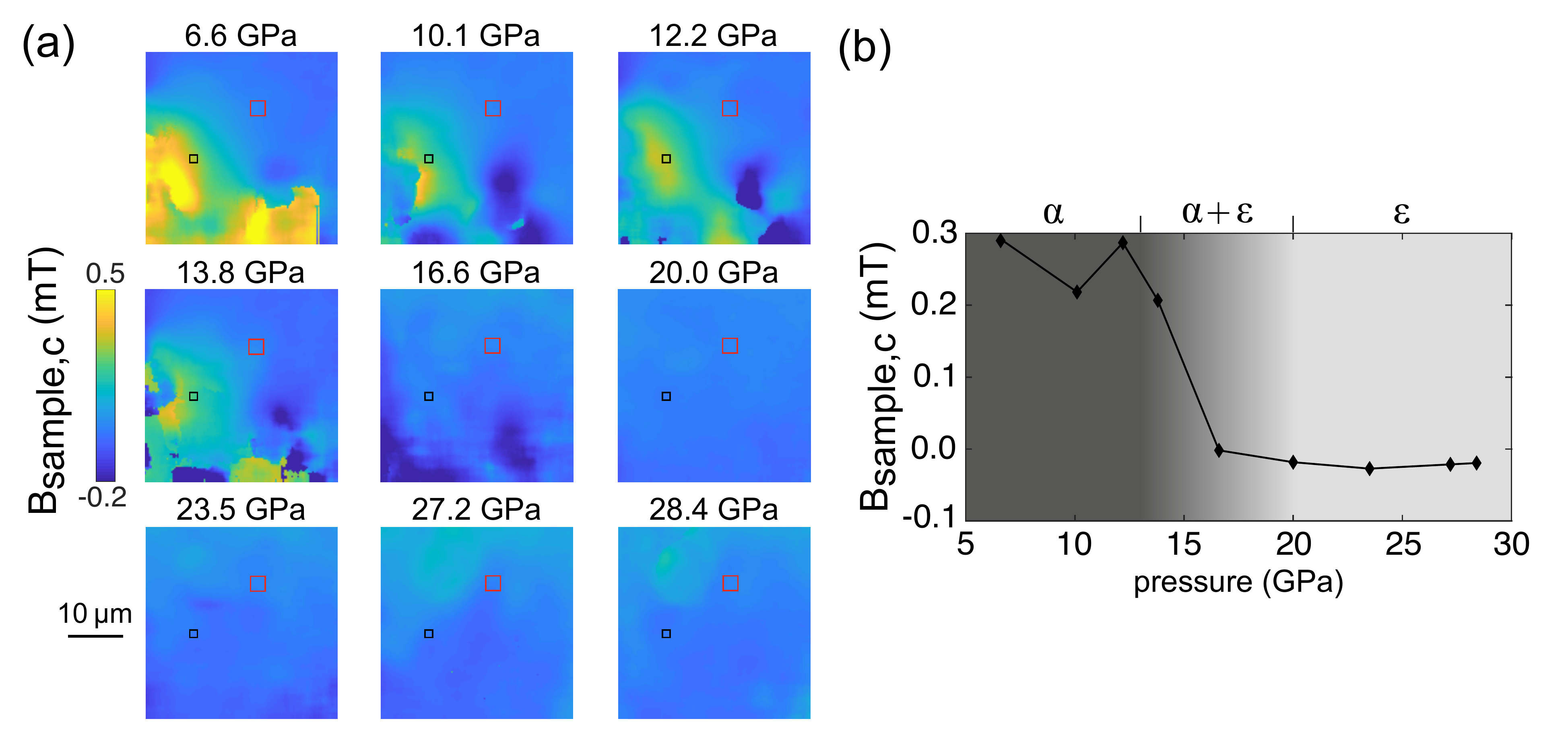}
 \caption{\textbf{(a)} Color maps of the magnetic field projection $B_{{\rm sample},{\mathrm{C}}}$ of the magnetic field created by the iron magnetization  along the $c$ axis of the selected NV family. As the pressure increases, the local positive and negative variations of $B_{{\rm sample},{\mathrm{C}}}$ progressively disappear, until the sample no longer exhibits any magnetic properties. Red squares ($7 \times 7$ pixels) indicate the chosen reference area to extract sample magnetization effects from pressure and bias field ones. Black squares ($11 \times 11$ pixels) indicate the area over the sample where signal is averaged for figure 2b. \textbf{(b)} Measurement of $B_{{\rm sample},{\mathrm{C}}}$ averaged over the black square shown in panel (a) as the pressure in the DAC chamber increases.  A greyscale pattern is overlaid to indicate the start and the end of the \textalpha-Fe to  \textepsilon-Fe structural transition independently  determined using the XRD pattern. The evolution of the   magnetic field created by the iron magnetization is consistent with the XRD data.}
 \label{figMap}
 \end{figure}

Complementary to the XRD measurement, we analyzed the maps of the splitting value $\Delta_{\mathrm{C}}$ from the  NV-ODMR spectrum which were determined for each pressure step. As previously described~\cite{barson2017,broadway2019}, hydrostatic pressure induces a shift of the center frequency of the  spin resonance  while non-hydrostatic stress components quadratically add to the frequency splitting induced by the magnetic field. Therefore the analysis of the NV-ODMR spectrum under pressure requires to discriminate between the influence of stress on the NV center spin states and their response to the applied magnetic field. In order to retrieve the relevant splitting associated to the magnetic field created by the iron magnetization, we used reference values of the splitting at a distance of 15~\textmu m from the sample. On such length scale  the external magnetic field can be considered as uniform, and stress induced variations measured were below 0.2~GPa (2~MHz ODMR frequency shift at most). Moving this reference zone, highlighted as a red square in figure~\ref{figMap}.a, in the vicinity of the sample changed the magnetic field estimation results only by a factor smaller than other error sources. This procedure allowed us to remove both the contribution of the non-hydrostatic components of the stress tensor and the contribution of the applied magnetic field~\cite{lesik2019}. We could then extract the projection along the chosen $c$ axis
 of the magnetic field that is only created by the  magnetization of the samples located within the studied area of the pressure chamber (figure~\ref{figMap}.a). Since the vector magnetization of the iron samples remains aligned along the  applied magnetic field, the projection along the fixed $c$ axis is a good estimator of the  amplitude  of the magnetic field that is created by the samples. 
At low pressure below $10\, {\rm GPa}$  a   magnetic field  of a few hundreds of~\textmu T appeared above the iron samples due to the \textalpha-Fe magnetization. As the pressure increased, this magnetic field  decreased and finally disappeared. The data show that the magnetic transition starts between 12.2~GPa and 13.8~GPa, and ends at a pressure slightly above 20.0~GPa. Figure~\ref{figMap}.b gives a comparison between the magnetic transition inferred from the NV-ODMR signal and the structural transition determined by the XRD pattern. From these results, we confirm that both transitions occur within a similar pressure range. 


To conclude, this experiment demonstrates that NV diamond magnetic microscopy is a robust magnetic measurement  technique which can be operated on a synchrotron beamline with the intrinsic constraints of this environment \cite{fu2020}.  As a proof-of-principle experiment, we followed simultaneously the evolution of the magnetization of an iron sample along the \textalpha-Fe to \textepsilon-Fe transition with the XRD pattern associated to these two phases. This combined set-up can be easily adapted depending on the investigated samples, for instance by optimizing the focusing of the X-ray beam on the sample inside the DAC. Our work demonstrates that NV diamond magnetic microscopy is well suited to complement fourth-generation synchrotrons whose improved brightness and focusing will enable the observation of the dynamic, electronic and structural properties of  materials with upcoming spatial resolution of about 100~nm.  
 
 \ack 
We acknowledge the synchrotron SOLEIL for provision of beamtime under Proposal No 20181196. The experiment was completed on the PSICH\'E beamline.
This project  received funding from the European Union's research and innovation programme (H2020-FETFLAG-2018-2020) under grant agreement No 820394 (ASTERIQS), the  Agence Nationale de la Recherche (project SADAHPT),  the Investissements d'Avenir program through  LabEx PALM (ANR-10-LABX-0039-PALM),  and the DIM SIRTEQ of  R\' egion  \^{I}le-de-France. 
B. Vindolet acknowledges support by a PhD research grant of D\'el\'egation G\'en\'erale de l'Armement.  

\newpage

\section*{References}

\bibliographystyle{MyThesis}
\bibliography{BibSoleilNV}

\newpage

\end{document}